\documentclass[10pt]{article}
\usepackage{geometry}
\usepackage{amsfonts,amsmath,amsthm,amssymb,cancel,framed, color}
\usepackage[normalem]{ulem}
\usepackage[colorlinks=true, pdfstartview=FitV, linkcolor=black, citecolor=black, urlcolor=blue]{hyperref}

\definecolor{shadecolor}{rgb}{0.98, 0.98, 0.9}

\usepackage{tikz, mathrsfs}
\usepackage{tikz, pgfplots, todonotes,ulem,amssymb}
\usetikzlibrary{decorations.markings,arrows, calc, shapes, intersections, patterns}
\tikzstyle directed=[postaction={decorate,decoration={markings,
		mark=at position .65 with {\arrow{latex}}}}]

\def\dd{{\rm d}}

\def \wh {\widehat}
\def \N {\mathbb N}

\def\Fcal{{\mathcal F}}

\def\le{\left}
\def\ri{\right}
\def \QED{\hfill $\blacksquare$\par \vskip 4pt}
\def \bea#1\eea {\begin{align} #1 \end{align}}
\def \wt{ \widetilde }

\newtheorem{theorem}{Theorem}[section]
\newtheorem{proposition}{Proposition}[section]
\newtheorem{corollary}{Corollary}[section]
\newtheorem{remark}{Remark}[section]
\newtheorem{lemma}{Lemma}[section]

\def\nn{\nonumber}
\def\be{\begin{equation}}
\def\ee{\end{equation}}
\def\ba{\begin{array}}
\def\ea{\end{array}}
\makeatletter
\@addtoreset{equation}{section}
\makeatother

\def\Acal{\mathcal A}
\def \eqref #1{(\ref{#1})}
\def \1{\mathbf 1}
\def \br{\begin{remark}}
\def \er{\end{remark}}

\def\Z{{\mathbb  Z}}
\def\a{\alpha}
\def\s{\sigma}
\def\p{\partial}
\def\Mcal{{\mathcal M}}
\def\f{\frac}
\def\la{\label}
\def\tr{{\rm tr}}

\def\la{\label}

 \title{On the tau function of the hypergeometric equation}
\begin{document}
\begin{center}
\begin{Large}
On the tau function of the hypergeometric equation
\end{Large}

\bigskip
M. Bertola$^{\dagger\ddagger\diamondsuit}$\footnote{Marco.Bertola@concordia.ca, mbertola@sissa.it},  
D. Korotkin$^{\dagger\ddagger}$ \footnote{Dmitry.Korotkin@concordia.ca},
\\
\bigskip
\begin{small}
$^{\dagger}$ {\it   Department of Mathematics and
Statistics, Concordia University\\ 1455 de Maisonneuve W., Montr\'eal, Qu\'ebec,
Canada H3G 1M8} \\
\smallskip
$^{\ddagger}$ {\it  Centre de recherches math\'ematiques,
Universit\'e de Montr\'eal\\ C.~P.~6128, succ. centre ville, Montr\'eal,
Qu\'ebec, Canada H3C 3J7} \\
\smallskip
$^{\diamondsuit}$ {\it  SISSA/ISAS,  Area of Mathematics\\ via Bonomea 265, 34136 Trieste, Italy }\\
\end{small}
\end{center}
\vspace{0.5cm}

{\bf Abstract.} The monodromy map for a  rank-two system of differential equations with three Fuchsian singularities is classically solved by the Kummer formul\ae\ for Gauss' hypergeometric functions. 
We define the tau-function of such a system as the
generating function of the extended monodromy symplectomorphism, using an idea recently developed. This formulation allows us to  determine the dependence of the tau-function  on the monodromy data. Using the explicit solution of the monodromy problem, the tau-function is then explicitly written in terms of Barnes $G$-function. 
 In particular, if the Fuchsian singularities are placed to $0$, $1$
and $\infty$, this gives the structure constants of the 
asymptotical  formula of Iorgov-Gamayun-Lisovyy  for solutions of Painlev\'e VI equation. 

\tableofcontents

\section{Introduction}

The theory of isomonodromic deformations  and Painlev\'e equations is a classical subject and yet it has seen a  great boost of activity during the last  30 years. In the 1980's works of Jimbo, Miwa and their collaborators the isomonodromic deformations were considered from the point of view of quantum field theory and the fundamental notion of the isomonodromic tau-function $\tau$ was introduced \cite{JMU1, JMU2}. In particular, the divisor of zeros of the tau-function determines the locus of non-solvability of the inverse monodromy problem \cite{Miwa,Palmer:Zeros}.  Originally the isomonodromic tau-function was defined only as a function of the isomonodromic times while its dependence on the monodromy data was essentially overlooked. 
Recently the consistent way of fixing the monodromy dependence was proposed in \cite{Bertola08, CMP} where the tau-function was interpreted as the 
generating function of the monodromy symplectomorphism, with the Poisson structure on the connection side given by the Kirillov-Kostant bracket and the Poisson structure on monodromy side given by Goldman's bracket and their natural extensions. 
In the framework of \cite{CMP} the symplectic potential on the monodromy manifold was defined with the help of   Fock-Goncharov coordinates
and equations for $\tau$ with respect to these coordinates were derived.

An important recent development in the theory of the tau-function of the Painlev'e VI equation was the discovery of Gamayun-Iorgov-Lisovyy 
(the so-called ``Kiev formula" \cite{Kiev, Kiev2}) which uncovered  the meaning  of the coefficients of the asymptotic expansion of the tau-function.
The structure constants of the Kiev formula are related to the  three-point  correlation function of the Liouville theory \cite{DO,DO1,ZZ}  when the central charge tends to $1$. Difference equations for the structure constants are special case of equations   found in a 1995 paper by Teschner \cite{Teschner}.

In this paper we apply the formalism developed in   \cite{CMP} to the seemingly trivial case of the Fuchsian system with three regular singularities. 
The dependence on the isomonodromic times  is explicit; however, the  tau-function has a non-trivial dependence on the monodromy data. Remarkably, the essential part of the tau-function understood as the generating function of the monodromy map, coincides in this case with the  structure constants of the Kiev formula \cite{Kiev, Kiev2}.


Let us now discuss our results in greater detail.
We start from a general   $sl(2)$  system with three Fuchsian singularities
\be
\f{d\Psi}{d z}=\left(\frac{A_0}{z-t_0}+\frac{A_1}{z-t_1}+\frac{A_\infty}{z-t_\infty}\right)\Psi\;,
\la{eq}
\ee
where $A_0+A_1+A_\infty=0$ and $\Psi$ is $2\times 2$ matrix. The theory is rather elementary: indeed, if we denote the eigenvalues of  $A_0$, $A_1$ and $A_\infty$ by $\pm \theta_0$, $\pm \theta_1$ and $\pm \theta_\infty$, respectively, then the matrix ODE can be reduced to the classical Gauss' hypergeometric equation.
However, it turns out that the theory of equation (\ref{eq}), being embedded appropriately in the theory of Fuchsian equations with an
arbitrary number of singularities is rather non-trivial.  Namely, the  equation (\ref{eq}) is the simplest  example where we can apply  the theory of extended monodromy symplectomorphism for Fuchsian systems developed in the recent paper \cite{CMP}.
According to the theory of \cite{CMP}, the generating function of the symplectomorphism can  be interpreted as the isomonodromic Jimbo-Miwa tau-function, and this allows to determine in a natural way its dependence on the  monodromy data (the original Jimbo-Miwa tau-function is defined only up to an arbitrary monodromy-dependent factor).
 The dependence of the tau-function on the positions $t_0$, $t_1$ and $t_\infty$ of the poles is defined by the isomonodromic equations of \cite{JMU1}
\be
\f{\p\log\tau}{\p t_j}=H_j , \ \qquad H_j 
=  \sum_{k:\  k\neq j} {\rm tr}\f{A_k A_j}{t_j-t_k}
\la{JM}\ee
with $j,k=0,1,\infty$. This dependence is seen to be rather simple: namely, since $A_0+A_1+A_\infty=0$, we have 
 $$\tr A_0 A_1 = \theta_\infty^2 -\theta_0^2-\theta_1^2\hskip0.7cm {\rm etc.}$$ 
Then equations  (\ref{JM}) imply 
\be
\tau=\tau_0(\theta_0,\theta_1,\theta_\infty) \,(t_0-t_1)^{ \theta_\infty^2-\theta_0^2-\theta^2_1 } (t_0-t_\infty)^{\theta^2_1-\theta^2_\infty-\theta^2_0} (t_1-t_\infty)^{ \theta^2_0-\theta^2_1-\theta^2_\infty}\;.
\la{tauz}
\ee
for some $t_j$-independent prefactor $\tau_0$. \\[3pt]

The prefactor $\tau_0$ is the focus of this short note: we want to fully  characterize
 its  dependence on the  monodromy data $\theta_0$, $\theta_1$ and $\theta_\infty$.
  This dependence is determined according to  \cite{CMP}
by requiring $\tau$ to be the generating function of the natural symplectomorphism given by the extended monodromy map which is, loosely speaking,  the map between the set of eigenvectors of matrices $A_j$ and the set of eigenvectors of the monodromy matrices $M_j$, see  (\ref{taumon}) for details.

This gives the following formula for $\tau_0$:
\be
\tau_0(\theta_0,\theta_1,\theta_\infty)=\frac{ G(1+\theta_0 + \theta_1+\theta_\infty  )
G(1-\theta_0+\theta_1+\theta_\infty)
G(1+ \theta_0 - \theta_1 + \theta_\infty)
G(1+  \theta_0 + \theta_1 - \theta_\infty)}
{G(1+ 2\theta_0) G(1+2\theta_1) G(1+2\theta_\infty)}
\la{DOZZ}
\ee
where $G$ is the Barnes' $G$-function.

The function $\tau_0$ appears as a structure constant of the asymptotic formula for solutions of Painlev\'e VI equation
\cite{Kiev,Kiev2}. It it also related to the computation of the topological string partition function  \cite{trinion}.


 Along similar lines we arrive to a new interpretation of  the difference equations for $\tau_0$ derived by Teschner in \cite{Teschner} in terms of Euler's $\Gamma$ function, before the Zamolodchikovs' formula became available:
\be
\f{\tau_0(\theta_0+1,\theta_1,\theta_\infty)}{\tau_0(\theta_0,\theta_1,\theta_\infty)}=
\frac{
 \Gamma(1-\theta_0 + \theta_1 + \theta_\infty) \Gamma(2\theta_0) \Gamma(2\theta_0-1)}
 {\Gamma(\theta_0 -\theta_1 + \theta_\infty) \Gamma(\theta_0 + \theta_1- \theta_\infty) \Gamma(\theta_0 + \theta_1 + \theta_\infty)}.
 \la{Teq}
\ee

In our context, this difference equation arises as the result of the standard (discrete) Schlesinger transformation \cite{JMU2} to the system (\ref{eq}). The expression \eqref{Teq} immediately follows from \eqref{DOZZ} due to  the functional equation of the Barnes' $G$-function; however,   historically \eqref{Teq} where derived first (and moreover in a more general setting of conformal field theory with arbitrary central charge, while the above formula is the specialization to unit central charge) and then used to get the explicit formula for the three-point conformal block. Here we observe that this formula  has a clear origin within the theory of the isomonodromy (discrete) transformations  \cite{JMU2}.

\section{Isomonodromic tau-function as generating function of extended monodromy symplectomorphism}
\la{monmap}

In this section we briefly review the extended monodromy map.
Let $\Psi(z)$ denote a solution of the ODE \eqref{eq} on the simply connected domain obtained by removing from the Riemann sphere the  two segments joining the three points $t_0,t_1,t_\infty$. It is known that near the Fuchsian singularities the matrix $\Psi(z)$ admits the following local representation 
\be
\Psi(z)=G_j\big(\1 + \mathcal O(z-t_j)\big)(z-t_j)^{L_j} C_j^{-1}
\la{as}
\ee
where the power is understood with a branch-cut along the chosen segments and the analytic prefactor has radius of convergence equal to the distance from $t_j$ or the nearest other singularity. 

The matrix $C_j$ is called the {\it connection matrix};  we can call the matrix $G_j$ {\it eigenvector matrix}. The (diagonal)  matrix $L_j=\theta_j \sigma_3$ is called {\it exponent of monodromy} and it consists of the eigenvalues  of the residue matrices $A_j$ which we assume to satisfy the non-resonance condition  $\theta_j \not\in \frac 12 \Z$:
\be
A_j=G_j L_j G_j^{-1}\;,\hskip0.7cm \sum_{j=0,1,\infty} A_j=0.
\la{AG}
\ee
The normalization of $\Psi(z)$ can be chosen in many ways; typically one chooses it so that one of the connection matrices  is the identity matrix. 

Corresponding to this setup, analytic continuation of $\Psi(z)$ along counterclockwise paths yields 
the monodromy matrices $M_0$, $M_1$ and $M_\infty$ of equation (\ref{eq}) around the poles $t_0$, $t_1$ and $t_\infty$  respectively; we assume that they are ordered such that $M_1 M_0 M_\infty =I$.    

 Similarly to the matrices $G_j$, the  matrices $C_j$ consist of the eigenvectors  of the  monodromy matrices:
\be
M_j= C_j \Lambda_j C_j^{-1}\qquad \Lambda_j:={\rm e}^{2\pi i L_j}.
\la{MC}
\ee
Let us briefly review the results of the paper \cite{CMP}, valid for any number $N$ of poles $t_j$ (and any rank).
The extended monodromy map $\Fcal^t$ sends the set  
{
\be
\label{Acal}
\Acal=\bigg\{(G_j,L_j)_{j=1}^N:\ \ \sum_{j=1}^N G_jL_j G_j^{-1}=0\bigg\}/\sim
\ee
 where the quotient is by the simultaneous transformation  $G_j\to S G_j$ (which corresponds to simultaneous conjugations of the matrices $A_j$)   to the set 
\be
\label{Mcal}
\Mcal=\bigg\{(C_j,L_j)_{j=1}^N:\ \ \prod_{j=1}^N C_j{\rm e}^{2i\pi L_j} C_j^{-1}=\1\bigg\}/\sim
\ee
where the quotient is, again, by the relation $C_j \to S' C_j$ (corresponding to the simultaneous conjugation of all monodromy matrices). 
In contrast,  the regular monodromy map sends the set $\{A_j\}/\sim $ to the set $\{M_j\}/\sim$.
}
This map is known to be a local (analytic)  diffeomorphism;
the complement of its image in $\Mcal \times\mathcal T$ (with $\mathcal T := \{t_j:\  t_j\neq t_k,\ j\neq k\}_{j=1}^N $) is  a co-dimension $1$ subvariety generally referred to as the {\it Malgrange divisor}, and we denote it by $\Delta_{\text{Mal}}$ .  The inverse monodromy map, conversely, is also an analytic local diffeomorphism defined on $\Mcal \times \mathcal T\setminus \Delta_{\text{Mal}}$; it is known that the inverse monodromy map has a  pole on $\Delta_{\text{Mal}}$. Our considerations are  of local nature and we stay away from the Malgrange divisor, we can  consider the data $(G_j,L_j,t_j)_{j=1}^N$ as function of the data $(C_j, L_j, t_j)_{j=1}^N$, or viceversa, as the context requires.

The natural symplectic form on the space $\Acal$ is given by the expression \cite{CMP,BBT}:
 \be
 \omega_\Acal=\sum_{k=1}^N\tr (\dd L_k \wedge G_k^{-1} \dd G_k)-\sum_{k=1}^N\tr ( L_k  G_k^{-1} \dd G_k\wedge G_k^{-1} \dd G_k)\;
 \la{OA}
 \ee
 where $\dd$ denotes the differential with respect to monodromy parameters. The Poisson bracket between $G_j$'s and $L_j$'s induced by the symplectic form (\ref{OA}) are quadratic; they are defined by the dynamical $r$-matrix (see \cite{CMP}).
The symplectic form can be expressed as  $\omega_\Acal=\dd\Theta_\Acal$ and the {\it symplectic potential} $\Theta_\Acal$ is given by 
\be
  \Theta_\Acal=\sum_{k=1}^N\tr (L_k G_k^{-1} \dd G_k)\;
 \la{thA}\ee
 where we emphasize that  $\dd$ denotes here the full differential on the space  of data $\{(C_j, L_j,t_j)\}_{j=1}^N$  and we view the matrices $G_j$'s as {\it functions} of the extended monodromy data (and the position of the poles) via the inverse monodromy map.
The natural symplectic form on the space $\Mcal$ is given by the following formula \cite{CMP}, which is a natural extension of
the Alekseev-Malkin expression (see (3.14) of \cite{AlekMal} in the case $g=0$):
\be
4\pi i \omega_{\Mcal}=
 \tr\sum_{\ell=1}^{N} 
\left( M_\ell^{-1} \dd M _{\ell} \wedge K_{\ell }^{-1} \dd K_{\ell} \right) 
+\tr  \sum_{\ell=1}^N  \left( \Lambda^{-1}_\ell C_{\ell} ^{-1}\dd C_{\ell}\wedge  \Lambda_\ell  C_\ell ^{-1}\dd C_\ell \right)
+2 \tr \sum_{\ell=1}^N  \left(\Lambda_\ell^{-1} \dd \Lambda_\ell \wedge C_\ell^{-1} \dd C_\ell
\right)
\la{OM}
\ee
with  $K_{\ell}=M_1\cdots M_\ell$ and $\Lambda_j=e^{2\pi i L_j}$.

The statement proven in \cite{CMP} is that the map $\Fcal^t$ is a symplectomorphism i.e. 
\be
(\Fcal^t)^* \omega_\Mcal=  \omega_\Acal\;.
\la{symint}
\ee
For ease of notation we will omit henceforth the explicit mention of the pull-back since we take the point of view that the matrices $C_j$ or $G_j$ are just different coordinates on the same manifold.
As it was suggested in \cite{CMP}, the natural definition of the tau-function which includes its dependence on times $t_j$'s and monodromy data, consists of 
\be
\dd\log \tau= \Theta_\Acal-\Theta_\Mcal - \sum_{j=1}^{N} H_j \dd t_j
\la{taumon}
\ee
where $\Theta_\Mcal$ is a potential for the form (\ref{OM}), which for an arbitrary $N$ was constructed in \cite{CMP} using the complex shear (or Fock-Goncharov) coordinates on the monodromy manifold. 

The dependence of $\tau$ on $t_j$ for fixed monodromy data then implies the original Jimbo-Miwa equations (\ref{JM}) (notice that 
$\Theta_\Acal $, being restricted to the subspace of constant  monodromy data, equals to $2\sum_{j=1}^{N} H_j \dd t_j$, see \cite{CMP}, Thm. 3.1).

\begin{remark}\rm
The formula \eqref{taumon} has an intrinsic ambiguity: while the potential $\Theta_\Acal$ is uniquely and globally defined \eqref{thA},  the potential $\Theta_\Mcal$  is not. This feature manifests the fact that the $\tau$ function is actually  a section of a certain (possibly projective) line bundle when the dependence on the monodromy data is taken into account. This was indicated in \cite{Bertola08} and then further investigated in \cite{CMP}. 

This is clearly apparent, even in this almost trivial setting, from \eqref{tauz}; indeed the $\tau$ function is a {\it multi-valued} function on $\Mcal\times \mathcal T$,  the multi-valuedness occurs only around the discriminant divisor $t_i=t_j$. 

We observe that if all {\it squares} of the exponents are integers, $\theta_j^2\in \Z$ then the $\tau$ function is actually a  globally defined (meromorphic) function of $t_j$'s.
\end{remark}

In the present case $N=3$ the form $\omega_\Mcal$ is especially simple, which allows us to find the potential 
$\Theta_\Mcal$ right away and then compute the Jimbo-Miwa tau-function $\tau$ explicitly.  

\section{Connection problem for three-point Fuchsian system and hypergeometric function}


We start from the following technical result expressing explicitly the connection matrices $C_j$, the  corresponding eigenvector matrices $G_j$ and the residue matrices $A_j$ for $N=3$ in terms of $\Gamma$-functions.

It is a rather straightforward computation which, however, seems to be absent from the literature; indeed the classical literature is more focused on the forward monodromy problem, namely, that of finding the connection formul\ae for given values of the parameters. We are rather interested in the inverse monodromy problem, whereby the connection matrices are explicit simple functions of the parameters, and the coefficient of the equation contain the transcendental part (which involves Gamma functions).

Consider the Fuchsian equation for the function $\Psi(w)$:
 \be
\label{HyperODE}
\frac {{\rm d}}{{\rm d} w}\Psi= \left(\frac {A_0}w + \frac {A_1}{w-1}\right)\Psi
\ee
where
\be
A_0
=  \left[ \begin {array}{cc} \theta_{0}&0\\ \noalign{\medskip}0&-\theta_
{0}\end {array} \right] \;,
\la{A0}
\ee
\be
A_1 =   
 \left[ \begin {array}{cc} 
 {\frac {  {\theta_{\infty}}^{2} -{\theta_{0}}^{2}-{\theta_{1}}
^{2}}{2\theta_{0}}}
&
{\frac {
\Gamma
 \left( \theta_{\infty}+\theta_{1}-\theta_{0}+1 \right)  \left( \Gamma
 \left( 2\,\theta_{0} \right)  \right) ^{2}}{
{\Gamma\le(\theta_0 - \theta_1+\theta_\infty\ri)}
 \Gamma \left( \theta
_{0}+\theta_{1}-\theta_{\infty} \right) \Gamma \left( \theta_{\infty}+
\theta_{1}+\theta_{0} \right) }}
\\
 \noalign{\medskip}
 {\frac {
 {\Gamma(\theta_0 - \theta_1+\theta_\infty+1)}
 \Gamma \left( \theta_{\infty}+\theta_{1}+\theta_{0}+1 \right) \Gamma
 \left( \theta_{0}+\theta_{1}-\theta_{\infty}+1 \right) }{
 \left( \Gamma \left( 2\,\theta_{0}+1 \right)  \right) ^{2}\Gamma
 \left( -\theta_{0}+\theta_{1}+\theta_{\infty} \right) 
 }}
&
  {\frac {  {\theta_{0}}^{2}+{\theta_{1}}^{2}- {\theta_{\infty}}^{2} }{2\theta_{0}}}
\end {array} \right]\;.
\la{A1}
\ee

Denote 
$$
A_\infty=-(A_0+A_1 ) 
$$
\be
= \left[ \begin {array}{cc} 
 {\frac { -{\theta_{0}}^{2}+{\theta_{1}}^{2}  -{\theta_{\infty}}^{2} }{2\theta_{0}}}
&
-{\frac {
\Gamma
 \left( \theta_{\infty}+\theta_{1}-\theta_{0}+1 \right)  \left( \Gamma
 \left( 2\,\theta_{0} \right)  \right) ^{2}}{
{\Gamma\le(\theta_0 - \theta_1+\theta_\infty\ri)}
 \Gamma \left( \theta
_{0}+\theta_{1}-\theta_{\infty} \right) \Gamma \left( \theta_{\infty}+
\theta_{1}+\theta_{0} \right) }}
\\
 \noalign{\medskip}
- {\frac {
 {\Gamma(\theta_0 - \theta_1+\theta_\infty+1)}
 \Gamma \left( \theta_{\infty}+\theta_{1}+\theta_{0}+1 \right) \Gamma
 \left( \theta_{0}+\theta_{1}-\theta_{\infty}+1 \right) }{
 \left( \Gamma \left( 2\,\theta_{0}+1 \right)  \right) ^{2}\Gamma
 \left( -\theta_{0}+\theta_{1}+\theta_{\infty} \right) 
 }}
&
  {\frac {  {\theta_{0}}^{2}-{\theta_{1}}^{2}+{\theta_{\infty}}^{2} }{2\theta_{0}}}
\end {array} \right]
\la{Ainf}
\ee
and
\be
q_j=e^{i\pi \theta_j}.
\la{qt}
\ee
Then we have  
$$A_j = G_j L_j G_j^{-1}$$ and 
$$M_j = C_j \Lambda_j C_j^{-1}$$
 (here $L_j=\theta_j \sigma_3$ and $\Lambda_j=e^{2\pi i L_j}$)  with the explicit expressions for $C_j$ and $G_j$ from the asymptotic expansion (\ref{as}) (with $t_0=0$, $t_1=1$ and $t_\infty=\infty$) given by the following proposition. 

\begin{proposition} \la{conn}

    
Consider the set of  connection matrices given by   
\be
C_0=\1,\ \ \ 
C_1 
=
 \left[ \begin {array}{cc} {\frac 
 {  q_0^2\le(q_{1}^{2}-q_{0}^{2}q_{\infty}^{2}\ri)}{q_{\infty}\,q_{1}\, \left( q_{0}^{4}-1 \right) }}
&
{\frac {q_{0}\, \left( q_{0}^{2}q_{1}^{2}q_{\infty}^{2}-1
 \right)  \left( q_{0}^{2}q_{1}^{2}-q_{\infty}^{2} \right) }
 {q_{\infty}^{2} \left( q_{1}^{4}-1 \right)  \left( q_{0}^{4}-1
 \right)   }
}\\ \noalign{\medskip}-\frac 1 { q_{0}} &{\frac 
{ \left(q_{0}^{2}-q_{\infty}^{2} q_{1}^{2} \right) q_{1}}{q_{\infty}\,q_{0}^{2} \left( {
q_{1}}^{4}-1 \right) }}\end {array} \right]
\la{C1}
\ee
\be
C_\infty
=
\left[ \begin {array}{cc} {\frac {iq_{0}\, \left( q_{0}^{2}q_{1}^
{2}-q_{\infty}^{2} \right) }
{q_{1}\,q_{\infty}^{2} \left( q_{0}^
{4}-1 \right)   }}&
{\frac {-i q_{\infty}\,
 \left( q_{0}^{2}q_{\infty}^{2}-q_{1}^{2} \right) 
  \left( q_{0}^{2}q_{1}^{2}q_{\infty}^{2}-1 \right) }
  {q_{1}^{2} 
\le(q_0^4-1\ri)\le(q_\infty^4-1\ri)
   }}
\\ \noalign{\medskip}
{\frac {-i}{q_{\infty}}}
&
{\frac {i \left( q_{0}^{2} -q_{\infty}^{2}q_{1}^{2}\right) q_{\infty}^{2}}{q_{1}\,q
_{0}\,  \left( q_{\infty}^{4}-1
 \right) }}\end {array} \right].
 \la{Ci}
\ee
Then the corresponding eigenvector matrices are  given by

$$
G_0=\1,\qquad G_1= \left[ \begin {array}{cc} -{\frac {\Gamma \left( \theta_{\infty}+
\theta_{1}-\theta_{0}+1 \right) \Gamma \left( 2\,\theta_{0} \right) }{
\Gamma \left( 2\,\theta_{1}+1 \right) \Gamma \left( \theta_{0}-\theta_
{1}+\theta_{\infty} \right) }}&{\frac {\Gamma \left( 2\,\theta_{0}
 \right) \Gamma \left( 2\,\theta_{1} \right) }{\Gamma \left( \theta_{0
}+\theta_{1}-\theta_{\infty} \right) \Gamma \left( \theta_{\infty}+
\theta_{1}+\theta_{0} \right) }}\\ \noalign{\medskip}-{\frac {\Gamma
 \left( \theta_{1}-\theta_{\infty}+1+\theta_{0} \right) \Gamma \left( 
\theta_{\infty}+\theta_{1}+\theta_{0}+1 \right) }{\Gamma \left( 2\,
\theta_{1}+1 \right) \Gamma \left( 1+2\,\theta_{0} \right) }}&-{\frac 
{\Gamma \left( 2\,\theta_{1} \right) \Gamma \left( \theta_{0}-\theta_{
1}+\theta_{\infty}+1 \right) }{\Gamma \left( 1+2\,\theta_{0} \right) 
\Gamma \left( -\theta_{0}+\theta_{1}+\theta_{\infty} \right) }}
\end {array} \right] 
$$
\be
G_\infty =  \left[ \begin {array}{cc} {\frac {i\Gamma \left( \theta_{\infty}+
\theta_{1}-\theta_{0}+1 \right) \Gamma \left( 2\,\theta_{0} \right) }{
\Gamma \left( \theta_{0}+\theta_{1}-\theta_{\infty} \right) \Gamma
 \left( 2\,\theta_{\infty}+1 \right) }}&{\frac {-i\Gamma \left( 2\,
\theta_{\infty} \right) \Gamma \left( 2\,\theta_{0} \right) }{\Gamma
 \left( \theta_{\infty}+\theta_{1}+\theta_{0} \right) \Gamma \left( 
\theta_{0}-\theta_{1}+\theta_{\infty} \right) }}\\ \noalign{\medskip}{
\frac {-i\Gamma \left( \theta_{\infty}+\theta_{1}+\theta_{0}+1
 \right) \Gamma \left( \theta_{0}-\theta_{1}+\theta_{\infty}+1
 \right) }{\Gamma \left( 1+2\,\theta_{0} \right) \Gamma \left( 2\,
\theta_{\infty}+1 \right) }}&{\frac {-i\Gamma \left( 2\,\theta_{\infty
} \right) \Gamma \left( \theta_{1}-\theta_{\infty}+1+\theta_{0}
 \right) }{\Gamma \left( 1+2\,\theta_{0} \right) \Gamma \left( -\theta
_{0}+\theta_{1}+\theta_{\infty} \right) }}\end {array} \right]
\la{Gjs}
\ee

%
%
%

\hfill $\blacktriangle$
\end{proposition}
\noindent{\bf Proof (sketch).}
The equation \eqref{HyperODE} can be reduced to the hypergeometric equation
\be\label{gauss}
w(1-w)\frac{{\dd}^{2}f}{{\dd w}^{2}}+\left(c-(a+b+1)w\right)\frac{\dd f}{\dd w}-ab f=0\;.
\ee
 with parameters
$$
a=\theta_{\infty}+\theta_{1}+\theta_{0},\ \ \ b=\theta_{1}-\theta_{
\infty} + \theta_{0}-1, \ \ \ c=2\,\theta_{0}\;.
$$
The explicit relationship is that our $\Psi(w)$ (with the connection matrix $C_0=\1$) is written as 
\bea
\label{Psiw}
\Psi(w)  &=(1-w)^{\frac{a+b-c+1}2} w^\frac c 2 \wt G_0^{-1}  H(w) \le[
 \begin{array}{cc}
 h_0(w) & s_0(w)\\\
 h_0'(w) & s_0'(w)
 \end{array}
 \ri] \a_0^{\frac {\s_3}2}
\eea
where $\a_0$ and $\wt G_0$ are appropriate constant expressions (w.r.t $w$) specified below in (\ref{G0t}), (\ref{a0}),  and
\bea
\label{Ghs}
&H(w) =  \left[ \begin {array}{cc} b&w-{\frac {b-c+1}{b-a+1}}, 
\\ \noalign{\medskip}0&{\frac {1}{ \left( 1-c \right) b}}\end {array}
 \right] \\ \ \ \ \  &h_{0}(w)={}_2{\rm F}_1\!\left({a,b\atop c};w\right),\ \ \ 
s_0(w) =w^{1-c} {}_2{\rm F}_1\!\left({a-c+1,b-c+1\atop 2-c};w\right).
\eea
From this expression and with aid of the classical Kummer formul\ae\ we can find the connection matrices as explained hereafter. 

We start from explaining the  pieces of the formula \eqref{Psiw};  denote 
\be
W(w) = (1-w)^{\frac{a+b-c+1}2} w^\frac c 2 \le[
 \begin{array}{cc}
 h_0(w) & s_0(w)\\\
 h_0'(w) & s_0'(w)
 \end{array}
 \ri].
\ee
Then $W$ 
 solves the system
\be
\frac {\dd W}{\dd w}= \le[
\begin{array}{cc}
\frac{(a+b+1)w-c}{2w(w-1)}  & 1\\
\frac {ab}{w(1-w)} & 
\frac{(a+b+1)w-c}{2w(1-w)}
\end{array}
\ri] W\;.
\ee
The gauge matrix $H(w)$ (\ref{Ghs}) is introduced to guarantee that the ODE  has poles only on the diagonal at $w=\infty$; indeed the matrix $\wt W(w):= H(w) W(w)$ satisfies the system 
\bea
\frac{\dd \wt W}{\dd w} =\left[ \begin {array}{cc}
{\frac {-w \left( b-a+1 \right) ^{2} -  \left( 2b-c+2 \right) a+c \left( b+1 \right) }{2 \left( -b+a-1 \right) w \left( w-1 \right) }}&
{\frac {b \left( 1-c \right) 
 \left( b+1 \right)  \left( b-c+1 \right)  \left( a-c \right) }{
 \left( b-a+1 \right) ^{2}w \left( w-1 \right) }}
\\ \noalign{\medskip}
{\frac {a}{ \left( c-1 \right) w \left( w-1
 \right) b}}&{\frac {w \left( -b+a-1 \right) ^{2}+ \left( 2\,b-c+
2 \right) a-c \left( b+1 \right) }{ 2\left( b-a+1 \right) w \left( 1-w
 \right) }}\end {array} \right]
\wt W
\label{Wtilde}
\eea
where the off-diagonal terms in the matrix of coefficients are now of order $\mathcal O(w^{-2})$  and  do not contribute to a pole at infinity. 

The remainder of the formula is a carefully crafted normalization which we now illustrate in some detail.

If we apply  Kummer formul\ae\ to find the connection matrices for the system \eqref{Wtilde} 
we obtain  matrices $C^{(0)}_{1,\infty}$ that contain expressions involving Gamma functions\footnote{The expression $w-\infty$ is a placeholder for $\frac 1 w$.};
\bea
\wt W(w) &= G_j^0 \Big(\1 + \mathcal O(w-j)\Big) (w-j)^{\theta_j\s_3} (C_j^{(0)})^{-1},\\
C_0^{(0)} := \1, \ \ \ 
C_1^{(0)}&:= \left[ \begin {array}{cc} {\frac {\Gamma \left( 1-c \right) \Gamma
 \left( a+b-c+1 \right)  \left( 1-c \right) }{\Gamma \left( a-c+1
 \right) \Gamma \left( b-c+1 \right) }}&{\frac {\Gamma \left( 1-c
 \right) \Gamma \left( c-a-b+1 \right) }{\Gamma \left( 1-a \right) 
\Gamma \left( 1-b \right)  \left( a+b-c \right) }}
\\ \noalign{\medskip}{\frac {\Gamma \left( c-1 \right) \Gamma \left( a
+b-c+1 \right)  \left( 1-c \right) }{\Gamma \left( a \right) \Gamma
 \left( b \right) }}&{\frac {\Gamma \left( c-1 \right) \Gamma \left( c
-a-b+1 \right) }{\Gamma \left( c-a \right) \Gamma \left( c-b \right) 
 \left( a+b-c \right) }}\end {array} \right]
\\
C_\infty^{(0)} &:=  \left[ \begin {array}{cc} {\frac {\Gamma \left( 1-c \right) \Gamma
 \left( a-b+1 \right) }{\Gamma \left( a-c+1 \right) \Gamma \left( 1-b
 \right)  \left( b-a \right) }}&-{\frac {\Gamma \left( 1-c \right) 
\Gamma \left( b-a+1 \right)  \left( c-1 \right) }{\Gamma \left( b-c+1
 \right) \Gamma \left( 1-a \right) {{\rm e}^{i\pi c}}}}
\\ \noalign{\medskip}-{\frac {{{\rm e}^{i \pi c}}\Gamma
 \left( c-1 \right) \Gamma \left( a-b+1 \right) }{\Gamma \left( a
 \right) \Gamma \left( c-b \right)  \left( b-a \right) }}&{\frac {
\Gamma \left( c-1 \right) \Gamma \left( b-a+1 \right)  \left( c-1
 \right) }{\Gamma \left( b \right) \Gamma \left( c-a \right) }}
\end {array} \right] 
 .
\eea
The matrices $G_j^0$ are instead given by 
\bea
G_0^0:= \left[ \begin {array}{cc} {\frac {b \left( b+1 \right)  \left( a-c
 \right) }{ \left( a-b-1 \right) c}}&-{\frac { \left( b-c+1 \right) 
 \left( -1+c \right) }{a-b-1}}\\ \noalign{\medskip} {\frac {a}{
 \left( 1 - c \right) c}}&{b}^{-1}\end {array} \right]
\  , \ \ \ \ \  &G_1^0 := \left[ \begin {array}{cc} {\frac {b \left( -1+c \right)  \left( b+1
 \right)  \left( b-c+1 \right) }{ \left( a-b-1 \right)  \left( a+b-c+1
 \right) }}&{\frac {a-c}{a-b-1}}\\ \noalign{\medskip}-{\frac {a}{a+b-c
+1}}&{\frac {1}{b \left( 1-c \right) }}\end {array} \right]
\\
&G_\infty^{0} = {\rm e}^{ \frac {i\pi(a-b+c-1)}2\s_3}.
\eea

However we prefer connection matrices that {\it do not contain} any Gamma function; the ``trick'' is to multiply $G_j^0$   on the right by suitable  diagonal matrices  $\a_j^{\frac {\s_3} 2}$, 
\be
G_j^{(0)} \to \wt G_j = G_j^{(0)} \a_j^{\frac {\s_3} 2}
\la{G0t}\ee   and simultaneously multiply the connection matrices with the same factors: $C^{(0)}_j \to \wt C_j =  C_j^{(0)}\a_j^{\frac {\s_3} 2}$. This operation is simply a convenient choice of normalizations for the eigenvectors of the matrices $A_j$ and the eigenvectors of the monodromy matrices. Inspection indicates that $\a_j$ can be chosen as follows:
\be
\a_0 = -{\frac {{{\rm e}^{-2\,i\pi\,\theta_{0}}}  \Gamma \left( \theta_{1}-\theta_{\infty}-1+\theta_{0}
 \right) \Gamma \left( \theta_{\infty}+\theta_{1}+\theta_{0} \right) 
\Gamma \left( \theta_{0}-\theta_{1}+\theta_{\infty}+1 \right) }{\Gamma
 \left( \theta_{\infty}+\theta_{1}-\theta_{0}+1 \right)  \Gamma
 \left( 2\,\theta_{0} \right) \Gamma(2\theta_0-1) }}\;,
 \la{a0}
\ee
\be
\a_1={\frac {\Gamma \left( \theta_{\infty}+\theta_{1}-\theta_{0}+1 \right) 
\Gamma \left( \theta_{1}-\theta_{\infty}-1+\theta_{0} \right) \Gamma
 \left( \theta_{\infty}+\theta_{1}+\theta_{0} \right) }{ \left( 1-2\,
\theta_{0} \right) \Gamma \left( \theta_{0}-\theta_{1}+\theta_{\infty}
+1 \right)  \left( \Gamma \left( 2\,\theta_{1} \right)  \right) ^{2}}}\;,
\la{a1}
\ee
\be
\a_\infty =
{\frac { \left(1- 2\,\theta_{0} \right) {{\rm e}^{-i\pi\,
 \left( 2\,\theta_{\infty}+2\,\theta_{0} \right) }}\Gamma \left( 
\theta_{\infty}+\theta_{1}-\theta_{0}+1 \right) \Gamma \left( \theta_{
0}-\theta_{1}+\theta_{\infty}+1 \right) \Gamma \left( \theta_{\infty}+
\theta_{1}+\theta_{0} \right) }{ \left( \Gamma \left( 2\,\theta_{
\infty}+1 \right)  \right) ^{2}\Gamma \left( \theta_{1}-\theta_{\infty
}-1+\theta_{0} \right) }}\;.
\la{ai}
\ee

To get to  the normalization $C_0=\1$ we now multiply $\wt W$ from the right by $(C_0 \a_0^{-\frac {\s_3}2})^{-1} = \a_0^{\frac {\s_3}2}$ (hence $\wt C_{1,\infty}$ on the {\it left} by $\a_0^{-\frac {\s_3} 2}$).
We thus obtain the final connection matrices $ C_j= \a_0^{-\frac{\s_3}2} C_{j}^{(0)} \a_j^{\frac {\s_3}2}$ which become, with the aid of Euler reflection formula, a trigonometric expression in $\theta_{0,1,\infty}$ and hence rational in $q_{0,1,\infty}$ (\ref{C1}), (\ref{Ci}) . 

At the same time we multiply $\wt W$ on the {\it left} by $\wt G_0^{-1}$, which amounts to a conjugation of the matrix of the ODE and produces the expressions (\ref{Gjs}) in the statement. \QED

\vskip 1cm

The connection and eigenvector matrices in Prop. \ref{conn} can still be acted upon by the gauge transformation that multiplies each pair  $G_j, C_j$ on the right by an arbitrary diagonal matrix ${\rm e}^{\rho_j\sigma_3}$. 
%
Namely,  the dimension of the extended character variety  is $\dim\Mcal=6$. 
%
%
%
The full set of local coordinates on $\Mcal$ is given by $(\theta_j,\rho_j)$, $j=0,1,\infty$.
We formalize this in the following corollary to Prop. \ref{conn}.
\begin{corollary}
For $t_0=0$, $t_1=1$ and $t_\infty=\infty$
the general solution of the inverse extended monodromy problem  is given by the connection matrices $C_j^\rho$  and  eigenvector   matrices $G_j^\rho$:
\be
C_j^\rho= C_j\left(\ba{cc} e^{\rho_j} & 0 \\ 0 & e^{-\rho_j}\ea\right)\;;
\la{Cjr}
\ee
\be
G_j^\rho= G_j\left(\ba{cc} e^{\rho_j} & 0 \\ 0 & e^{-\rho_j}\ea\right)
\la{Gjr}
\ee
where $C_j$'s are given by \eqref{C1}, \eqref{Ci} and $G_j$'s are given by (\ref{Gjs}).
\end{corollary}
\begin{remark}
For completeness we write here the form of the monodromy matrices:
\be
\label{Mon0}
M_0 =  \left[ \begin {array}{cc} q_{0}^{2}&0\\ \noalign{\medskip}0&q_{0}
^{-2}\end {array} \right]\;,
\ee
\vskip0.2cm
\be
\label{Mon1}
M_1 =  \left[ \begin {array}{cc} 
{\frac {q_{0}^{2}q_{1}^{2}q_{\infty}^
{4}- \left( q_{1}^{4}+1 \right) q_{\infty}^{2}+q_{0}^{2}q_{1}
^{2}}{ \left( q_{0}^{4}-1 \right) q_{1}^{2}q_{\infty}^{2}}}
&
{\frac {q_{0}^{3} \left( q_{0}^{2}q_{\infty}^{2}-q_{1}^{2}
 \right)  \left( q_{0}^{2}q_{1}^{2}-q_{\infty}^{2} \right) 
 \left( q_{0}^{2}q_{1}^{2}q_{\infty}^{2}-1 \right) }
 { \left( {q_
{0}}^{4}-1 \right) ^{2}q_{\infty}^{3}q_{1}^{3}}}\\ \noalign{\medskip}{\frac 
{q_{1}^{2}q_{\infty}^{2}-q_{0}^{2}}{q_{\infty}\,q_{0}^{3}q_{1}
}}&{\frac { \left( -q_{1}^{2}q_{\infty}^{4}+q_{0}^{2} \left( q_
{1}^{4}+1 \right) q_{\infty}^{2}-q_{1}^{2} \right) q_{0}^{2}}{
 \left( q_{0}^{4}-1 \right) q_{1}^{2}q_{\infty}^{2}}}
\end {array} \right] \;,
\ee
\vskip0.2cm
\be
\label{Monin}
M_\infty = \left[ \begin {array}{cc} {\frac {-q_{1}^{2}q_{\infty}^{4}+{q_{0}
}^{2} \left( q_{1}^{4}+1 \right) q_{\infty}^{2}-q_{1}^{2}}{
 \left( q_{0}^{4}-1 \right) q_{1}^{2}q_{\infty}^{2}}}&
 -{\frac {q_{0}\, \left( q_{0}^{2}q_{\infty}^{2}-q_{1}^{2} \right)  \left( 
q_{0}^{2}q_{1}^{2}-q_{\infty}^{2} \right)  \left( q_{0}^{2}{q_
{1}}^{2}q_{\infty}^{2}-1 \right) }{ \left( q_{0}^{4}-1 \right) ^{2
} q_{\infty}^
{3}q_{1}^{3}}}\\ 
\noalign{\medskip}{\frac {q_0^2 -q_{1}^{2}q_{\infty}^
{2}}{q_{0}\,q_{1}\,q_{\infty}}}&
{\frac { \left( q_{0}^{2
}q_{1}^{2}q_{\infty}^{4}- \left( q_{1}^{4} + 1 \right) q_{\infty
}^{2}+q_{0}^{2}q_{1}^{2} \right) q_{0}^{2}}{ \left( q_{0}^{4}
-1 \right) q_{1}^{2}q_{\infty}^{2}}}\end {array} \right] \;.
\ee
\end{remark}

\subsection{Connection problem for arbitrary $t_0$, $t_1$ and $t_\infty$}

If we move the poles from $0,1,\infty$ to an arbitrary triple $t_0,t_1,t_\infty$ with the aid of a M\"obius transformation, then  Proposition \ref{conn} leads to different eigenvector matrices as this proposition shows.
\begin{proposition}
Let $G_j^\rho$ and $C_j^\rho$ be given by (\ref{Cjr}), (\ref{Gjr}),  Prop. \ref{conn}. Then, under a conformal transformation sending $0,1,\infty$ to $t_0,t_1,t_\infty$,  the 
solution of the extended monodromy map for the same $C_j^\rho$ is given by the following set of eigenvector matrices $G_j^\rho(t_0,t_1,t_\infty)$:
\bea
G_0^\rho(t_0,t_1,t_\infty)&=G_0^\rho\cdot \left(w'( t_0)\right)^{L_0}=G_0^\rho \cdot\left(\f{t_1- t_\infty}{(t_0- t_\infty)(t_1- t_0)}\right)^{L_0}
\nn\\
G_1^\rho(t_0,t_1,t_\infty)&=G_1^\rho \cdot\left(w'( t_1)\right)^{L_1}=G_1^\rho \cdot \left(\frac{ t_\infty - t_0}{t_0-t_1)(t_1-t_\infty)} \ri)^{L_1}
\nn\\
G_\infty^\rho(t_0,t_1,t_\infty)&=G_\infty^\rho\cdot  \left((w^{-1})'( t_\infty)\right)^{L_\infty}=G_\infty^\rho\cdot\left(\f{t_1- t_0}{(t_\infty- t_0)(t_1- t_\infty)}\right)^{L_\infty}
\label{Gjst}
\eea
where the M\"obius transformation is given  by 
\be
w(z)=\f{(z- t_0)(t_1- t_\infty)}{(z- t_\infty)(t_1- t_0)}.
\ee
\end{proposition}
\noindent
{\bf Proof.}
The M\"obius map $w$ sends the points $t_0,t_1,t_\infty$ in the $z$--plane to $0,1,\infty$ in the $w$--plane. Under such a change of coordinate 
the solution of the ODE  transforms as a scalar $\wh \Psi(z) = \Psi(w(z))$. Let us denote $ w_0 = 0, w_1 =1$ and $w_\infty = \infty$;  we have 
\be
\wh \Psi(z) = {G_j^\rho}(t_0,t_1,t_\infty)\big(\1 + \mathcal O(z - t_j)\big)(z-t_j)^{L_j} 
\nn
\ee
\be
= G_j^\rho\Big(\1 + \mathcal O(z-w_j)\Big) \le( w(z) - w_j\ri)^{L_j} 
= G_j^\rho \Big(w'( t_j)\Big)^{L_j}\Big(\1 + \mathcal O(z-t_j)\Big) \le( z-  t_j\ri)^{L_j}.
\ee
In the above expression we understand the expression $w-\infty$ as $\frac 1 w$ (the local coordinate near $\infty$) so that $w'(t_\infty)$ is really $\frac \dd{\dd z} \le(\frac 1 {w(z)}\ri) \Big|_{z=t_\infty}$. This produces the expressions \eqref{Gjst}.\QED

\subsection{Three-pole isomonodromic tau-function}

To fix the dependence of $\tau$ on $t_j$'s we notice that, since  $A_0+A_1+A_\infty=0$, we have\footnote{Indeed, for example,  $\theta_\infty^2=\frac 12 \tr(A_\infty^2) = \frac 1 2 \tr( (A_0+A_1)^2) = \theta_0^2 + \theta_1^2 + \tr( A_0A_1)$.}
$$\tr A_0 A_1 = \theta_\infty^2 -\theta_0^2-\theta_1^2\;,\hskip0.7cm
\tr A_0 A_\infty = \theta_1^2 -\theta_0^2-\theta_\infty^2 \;, $$
$$\tr A_1 A_\infty = \theta_0^2 -\theta_1^2-\theta_\infty^2\;,$$
and, therefore,
 \be
 H_0=\f{\theta_\infty^2 -\theta_0^2-\theta_1^2}{t_0-t_1}+\f{\theta_1^2 -\theta_0^2-\theta_\infty^2}{t_0-t_\infty}\;,
 \hskip0.7cm
 H_1=\f{\theta_\infty^2 -\theta_0^2-\theta_1^2}{t_1-t_0}+\f{\theta_0^2 -\theta_1^2-\theta_\infty^2}{t_1-t_\infty}\;,
 \la{ham01}
 \ee
\be
 H_\infty=\f{\theta_1^2 -\theta_0^2-\theta_\infty^2}{t_\infty-t_0}+\f{\theta_0^2 -\theta_1^2-\theta_\infty^2}{t_\infty-t_1}\;.
 \la{hami}
\ee
It then follows from the  Jimbo-Miwa equations (\ref{JM}) that the $t_j$ dependence reads
\be
\tau
=\tau_0(\theta_0,\theta_1,\theta_\infty) \,(t_0-t_1)^{ \theta_\infty^2-\theta_0^2-\theta^2_1 } (t_0-t_\infty)^{\theta^2_1-\theta^2_\infty-\theta^2_0} (t_1-t_\infty)^{ \theta^2_0-\theta^2_1-\theta^2_\infty}\;.
\la{tauz2}
\ee
While the $t_j$ dependence is explicit and rather trivial, our interest is in the computation of the $t$--independent factor $\tau_0$.

Let us  make explicit the expressions involved in (\ref{taumon}) starting from $\Theta_\Mcal$.
The parametrization of the monodromy manifold in terms of $\theta_j$ and $\rho_j$ leads to the following lemma by a straightforward computation.
\begin{lemma}
The substitution of the matrices $C_j^\rho$ (\ref{Cjr}) and monodromies $M_j=C_j^\rho\Lambda_j (C_j^\rho)^{-1}= C_j\Lambda_j C_j^{-1}$ into (\ref{OM}) gives 
\be
\omega_\Mcal=2\sum_{j=0,1,\infty} \dd\theta_j \wedge\dd \rho_j \;.
\la{omtr}
\ee
\end{lemma}
The proof of the lemma is a straightforward but lengthy (Maple-assisted) computation.

The form (\ref{omtr}) is clearly symplectic; it is also closed and there are several obvious choices  of symplectic potential  such that 
$\dd \Theta_\Mcal=\omega_\Mcal$. The potential we use in this paper is
\be
\Theta_\Mcal=2\sum_{j=0,1,\infty} \theta_j\dd \rho_j\;,
\la{thetam}
\ee 
and differs slightly from the potential used in the general case in \cite{CMP} in terms of Fock-Goncharov coordinates due to a different choice of toric variables $\rho_j$.

Now  we are ready to formulate the main theorem of the paper.
%
\begin{shaded}
\begin{theorem}
\label{mandozz}
The tau-function $\tau$ defined by

\be
\dd\log \tau= \Theta_\Acal(L_j,G_j^\rho(\vec{t}))-\Theta_\Mcal (L_j,C^\rho_j) - \sum_{j=1}^{N} H_j \dd t_j,
\la{taumon1}
\ee
does not depend on $\rho_j$'s and is given by the following expression
\be
\tau=\tau_0(\theta_0,\theta_1,\theta_\infty) (t_0-t_1)^{\theta_\infty^2-\theta_0^2-\theta^2_1} (t_0-t_\infty)^{\theta^2_1-\theta^2_\infty-\theta^2_0} (t_1-t_\infty)^{\theta^2_0-\theta^2_1-\theta^2_\infty}\;
\la{tauz1}
\ee
with 
\be
\tau_0=\frac{ G(1+\theta_0 + \theta_1+\theta_\infty  )
G(1-\theta_0+\theta_1+\theta_\infty)
G(1+ \theta_0 - \theta_1 + \theta_\infty)
G(1+  \theta_0 + \theta_1 - \theta_\infty)}
{G(1+ 2\theta_0) G(1+2\theta_1) G(1+2\theta_\infty)}.
\la{DOZZ1}
\ee
\end{theorem}
\end{shaded}
\noindent {\bf Proof.} 
Observe that 
$$\Theta_{\Acal} (L_j,G^\rho_j(\vec t)) = \Theta_{\Acal} (L_j,G_j(\vec t)) + \sum_{j=0,1,\infty}2 \theta_j \dd \rho_j$$
 and the latter term precisely cancels $\Theta_\Mcal$. This shows that $\tau$ is independent of $\rho$ (even before an explicit computation). 
Let us  compute  the difference $\Theta_\Acal - \Theta_\Mcal$ in (\ref{taumon1})
 using the formulas of Prop. \ref{conn} together with the $t_j$ dependence of the eigenvector matrices given in  \eqref{Gjst}:
\be
\Theta_{\Acal} (L_j,G^\rho_j(\vec t))- \Theta_\Mcal 
=\sum_{j=0,1,\infty}\tr \Big( L_j (G_j^0)^{-1} \dd G_j^0\Big)
+  \sum_{j=0,1,\infty} \le[\tr \bigg(L_j  \dd\ln  T_j\bigg)  + 2\theta_j \dd \rho_j\ri]\nn
\ee
\be
+\sum_{j=0,1,\infty} 2 \theta_j \dd (\theta_j\ln(w'_j )) - 2\sum_{j=0,1,\infty} \theta_j\dd \rho_j
\la{thGL}
\ee
where we use the convention that $w'_\infty = (1/w)'(t_\infty)$ (due to the fact that the local coordinate near $w=\infty$ is $1/w$). 
Using \eqref{thetam} we observe that the term $2\sum \theta_j \dd \rho_j$ drops out. 

In this expression the only terms that depend on $\vec t$ are the in the last sum, while all the remaining terms can be  seen (using  (\ref{Gjs})) to depend only on $\vec \theta$ .

An elementary but long computations provides the  contribution of the first sum in \eqref{thGL}:
$$
\sum_{j=0,1,\infty} \tr \Big(L_j (G_j^0)^{-1} \dd G_j^0\Big) 
= \left( 2i\pi (\theta_{0}+\theta_{\infty})+ 
\frac 1{ \theta_{1
}-\theta_{\infty}-1+\theta_{0}}+{\frac {2\,\theta_{1}-2\,
\theta_{\infty}-1}{2\,\theta_{0}-1}} \right) \, {\rm d} \theta_0
$$
\be
+
 \left( 1+\frac 1{  \theta_{1}-\theta_{\infty}-1+\theta_{0}} \right) \, {\rm d} \theta_1+ \left( 2i\pi  \theta_{\infty}-1-
\frac 1{  \theta_{1}-\theta_{\infty}-1+\theta_{0} } \right) 
\, {\rm d} \theta_\infty \; .
\label{tt00}
\ee

For the second sum we have
\be
\tr \sum_{j=0,1,\infty} L_j  \dd \ln  T_j=\sum_{j=0,1,\infty} \theta_j \,\dd \ln \a_j
\la{ss}\ee
(with $\a_j$ given by (\ref{a0}),  (\ref{a1}), (\ref{ai})) which
gives a long expression involving $(\ln \Gamma)'$
and we are not going to write it down explicitly.

Therefore we get 
$$
\dd \log \tau = \Theta_{\Acal} (L_j,G^\rho_j(\vec{t})) -\Theta_\Mcal (L_j,C^\rho_j) -\sum_j H_j \dd t_j 
$$
\be
=\sum_{j=0,1,\infty}\tr \Big( L_j (G_j^0)^{-1} \dd G_j^0\Big) 
+ \sum_{j=0,1,\infty} \theta_j \,\dd \ln \a_j 
+\sum_{j=0,1,\infty} 2 \big(\theta_j \dd (\theta_j\ln(w'_j ))\big) - \sum_j H_j \dd t_j \;.
\la{taum2}
\ee

Let us now address  the dependence of $\tau$ on $\vec t$; this comes only from the second line in \eqref{taum2}; one can explicitly compute the derivatives of $w_j'$s with respect to $t_j$'s, use the expressions (\ref{ham01}), (\ref{hami}) for the Hamiltonians and verify that the result produces precisely the differential of the logarithm of the terms containing the $\vec t$--dependence in \eqref{tauz1}. 

The remainder is a  long but  straightforward computation that uses the explicit expression \eqref{tt00} and, most importantly the  expressions for the $\alpha_j$ contained in (\ref{a0}, \ref{a1}, \ref{ai}) for the term \eqref{ss}. 

It is helpful to use the following differential equation for the Barnes'  $G$--function:
\be
\label{BintG}
\ln G(z+1) =  \frac {z(1-z)}2 + \frac {z\ln(2\pi)}2 + z \ln \Gamma(z) - \int_0^z \ln \Gamma(\zeta)\dd \zeta\;.
\ee
Then, in the differential \eqref{ss} there will be many terms involving the expression $\ln \Gamma$, which can be integrated using \eqref{BintG}. 
Alternatively, one verify that the differential of (\ref{DOZZ1}) gives the explicit expression \eqref{taum2}.
\QED

\subsection{Teschner's equations and Schlesinger transformation}

The goal of this section is to show that  Teschner's equations \eqref{Teq}  can be interpreted as the action of  an elementary  Schlesinger transformation on the tau-function. Clearly,  the equation \eqref{Teq} can be immediately deduced from  the explicit formula \eqref{DOZZ}, so that this exercise is mostly of  pedagogical value. 

We start with a brief reminder of the definition of the discrete Schlesinger transformation, which was originally derived in \cite{JMU2} for arbitrary rational ODEs and generalized to arbitrary Riemann--Hilbert problems in \cite{Bertola08}.
For an arbitrary $2\times 2$ differential equation 
\be
\f{\dd \Psi}{\dd z}=\sum_{j=1}^N \f{A_j}{z-t_j} \Psi
\ee
with $A_j=C_j L_j C_j^{-1}$, $L_j=\theta_j \sigma_3$ and asymptotics
\be
\Psi(z)=G_j\big(\1 + \mathcal O(z-t_j)\big)(z-t_j)^{L_j} C_j^{-1}
\ee
a (discrete) Schlesinger transformations \cite{JMU2} consists in a suitable rational gauge transformation $A \to R^{-1} A R  -  R^{-1} R'$ crafted in such a way that the exponents of  monodromy are shifted by an integer and the monodromy matrices (as well as the connection matrices) remain the same. These are ``discrete'' isomonodromic equations and form a commutative lattice of transformations. We are interested in the sublattice that preserves the trace of $A$ (i.e. those for which the gauge matrix $R$ has unit determinant\footnote{In general $R(z)$ may have a rational determinant; but in this case it means that the resulting matrix $A$ will not be traceless.}). This sublattice has three generators that correspond to the shifts $\theta_j\mapsto \theta_j +1$. 
Such a generator is called an {\it  elementary Schlesinger transformation} and it maps $\Psi$ to the new function $\wt\Psi(z)=R_j(z)\Psi(z)$, according to the following lemma.
\begin{lemma}
Let $Y_j$ denote the holomorphic factor in the expansion of $\Psi$ near $t_j$;
\be
\label{345t}
\Psi(z) = Y_j(z) (z-t_j)^{L_j} C_j^{-1}, \ \ \ Y_j(z) = G_j + \mathcal O(z-t_j).
\ee
Define the matrix 
\be
R_j(z)= \1 - \frac {1} {\Big(Y_j^{-1}(t_j) Y_j'(t_j)\Big)_{12}} \frac { Y_j(t_j) \mathbb E_{21} Y_j^{-1}(t_j)}{z-t_j},\ \ \ j=1,\dots N.
\ee
where $\mathbb E_{21}$ is the elementary matrix.
Then the gauge transformation $\wt \Psi(z):= R_j(z) \Psi(z)$ preserves the extended  monodromy data and shifts $\theta_j \to \theta_j+1$: 
\be
\wt \Psi(z) =\wt G_j (\1 + \mathcal O(z-t_j)) (z-t_j)^{(\theta_j+1)\s_3} C_j^{-1}
\ee
\end{lemma}
These formulas are essentially contained in \cite{JMU2} but were generalized to arbitrary Riemann--Hilbert problems in \cite{Bertola08} (see Prop. 2.3 ibidem).

 \begin{figure}[t]
 \begin{center}
 \begin{tikzpicture}[scale=1.69]
\draw[line width = 1pt, postaction={decorate,decoration={{markings,mark=at position 0.7 with {\arrow[black,line width=1.5pt]{>}}}} }]
(0,0)--node[above, sloped]{{\resizebox{1.4cm}{!}{$ J(z) = M_1$}}}(15:1);

\draw[line width = 1pt, postaction={decorate,decoration={{markings,mark=at position 0.7 with {\arrow[black,line width=1.5pt]{>}}}} }]
(0,0)--node[above]{{\resizebox{1.4cm}{!}{$ J(z) = M_{_\ell}$}}}(180:1);

\draw[line width = 1pt, postaction={decorate,decoration={{markings,mark=at position 0.7 with {\arrow[black,line width=1.5pt]{>}}}} }]
(0,0)--node[above,sloped ]{{\resizebox{1.4cm}{!}{$ J(z) = M_{_N}$}}}(-60:1);
\draw [fill] circle [radius=0.03];
\node [below left] at (0,0) {$\infty$};
\foreach \x in {50,70,90,110,130,150, 220, 240,260}
\draw [fill] (\x:1.5)  circle [radius=0.01];

\draw [fill] (15:1)  circle[radius=0.02];
\draw [fill] (-60:1)  circle[radius=0.02];
\draw [fill] (180:1)  circle[radius=0.02];

\begin{scope}[shift={(15:1.5)}]
\draw[green!10!black,  thick,postaction={decorate,decoration={markings,mark=at position 0.9 with {\arrow[line width=1.5pt]{>}}}}] (0:0.5) arc (0:360:0.5) node [pos = 0.75, below,sloped] {\resizebox{2.5cm}{!}{$J(z) = C_{_1} (z-t_{_1})^{-L_{_1}}$}};

\node [right] at (0:0)  {$t_1$} ;
\draw [fill](0,0)  circle[radius=0.02];

\end{scope}

\begin{scope}[shift={(180:1.5)}]
\draw[green!10!black,  thick,postaction={decorate,decoration={markings,mark=at position 0.9 with {\arrow[line width=1.5pt]{>}}}}] (0:0.5) arc (0:360:0.5) node [pos = 0.75, below,sloped] {\resizebox{2.5cm}{!}{$J(z) = C_{_\ell} (z-t_{_\ell})^{-L_{_\ell}}$}};

\node [right] at (0:0)  {$t_{_\ell}$} ;
\draw [fill](0,0)  circle[radius=0.02];
\end{scope}

\begin{scope}[shift={(-60:1.5)}]
\draw[green!10!black, thick,postaction={decorate,decoration={markings,mark=at position 0.9 with {\arrow[line width=1.5pt]{>}}}}] (0:0.5) arc (0:360:0.5) node [pos = 0.75, below,sloped] {\resizebox{2.5cm}{!}{$J(z) = C_{_N} (z-t_{_N})^{-L_{_N}}$}};

\node [right] at (0:0)  {$t_{_N}$} ;
\draw [fill](0,0)  circle[radius=0.02];
\end{scope}

\end{tikzpicture}
\end{center}
\caption{Graph $\Sigma$ and jump matrices $J(z)$  on its edges used in the calculation of the form $\Theta_{\text{Mal}}$. On each smooth component of $\Sigma$ we indicate the corresponding jump matrix.}
\label{FigSchles}
\end{figure}
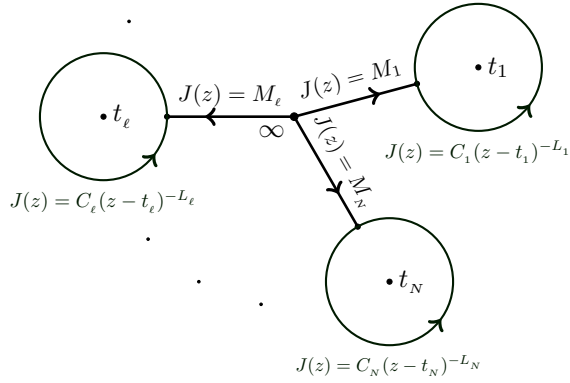

A Schlesinger transformation affects the eigenvector matrices and eigenvalues but not the monodromy matrices, and therefore it affects both the term $\Theta_\Acal$ and the Hamiltonians (albeit the latter in a simple way) entering the expression \eqref{taumon1}.
The effect is explained in the following proposition
\begin{proposition}
\label{propschles}
Under the elementary Schlesinger transformation $\theta_j \to \theta_j+1$ the tau function $\tau$ transforms to $\wt \tau$ according to 
\be
\frac {\wt \tau}{\tau} = \Big(Y_j^{-1}(t_j) Y'_j(t_j) \Big)_{12}
\label{346}
\ee
where $Y_j(z)$ is the local analytic expression \eqref{345t}.
\end{proposition}
\noindent {\bf Proof.}
Taking the differential of the logarithm of \eqref{346} and using the definition of $\tau$ in \eqref{taumon1}, the proposition reduces to the verification that 
\be
\wt \Theta_\Acal -\Theta_\Acal - \sum_{j=1}^N \Big(\wt H_j - H_j\Big) \dd t_j - \wt \Theta_\Mcal +\Theta_{\Mcal} = \dd \ln  \Big(Y_j^{-1}(t_j) Y'_j(t_j) \Big)_{12}.
\ee
We have observed in Theorem  \ref{mandozz} that $\tau$ is independent of the toric variables $\rho_j$'s and therefore can set them to $0$ and thus assume that $\Theta_\Mcal =\wt \Theta_\Mcal =0$ in this computation. 
In  \cite{CMP}, (Th.3.1) it was shown that  the combination
\be
\Theta_\Acal - \sum_{j=1}^N H_j \dd t_j = 
\Theta_{\text{Mal}}\;,
\ee
called the {\it Malgrange differential}, can be written as the following integral:
\be
\Theta_{\text{Mal}} = 
\int_\Sigma \tr \Big(\Psi^{-1}(z_-) \Psi'(z_-)\dd J(z) J(z)^{-1}\Big) \frac {\dd z}{2i\pi}\;.
\la{malf}
\ee 
The same statement (in slightly different notations) was derived in \cite{IP}, and called the {\it localization formula} there.
In (\ref{malf}) $\Sigma$ is the collection of contours shown in Fig.\ref{FigSchles} and $J(z)$ is the ``jump matrix''  on each arc of $\Sigma$ indicated in the same figure, and $\Psi(z_-)$ denotes value of the solution of the ODE evaluated on the left side of the arc at the point $z\in \Sigma$.  
The transformation of $\Theta_{\text{Mal}}$ under an elementary Schlesinger transformation was derived in Theorem 2.1 of \cite{Bertola08} and it is given (in the present notations)  by  
\be
\wt \Theta_{\text{Mal}} -\Theta_{\text{Mal}} = \dd \ln \Big(Y_j^{-1}(t_j) Y'_j(t_j) \Big)_{12} 
\ee
\QED

We specialize Proposition \ref{propschles} to the case $N=3$; in this case we have the explicit expression of the matrices $Y_j(z)$ in terms of hypergeometric functions, so that the ratio of tau functions under an elementary Schlesinger transformation can be explicitly computed with the formula \eqref{346}. As a result we get 
\bea
\frac {\tau(\theta_0+1, \theta_1,\theta_\infty)}{\tau(\theta_0,\theta_1,\theta_\infty)} =   
\frac {\tau_0(\theta_0+1, \theta_1,\theta_\infty)}{\tau_0(\theta_0,\theta_1,\theta_\infty)}
\left(\f{t_1-t_\infty}{(t_0-t_1)(t_0-t_\infty)}\right)^{2\theta_0+1} 
\eea
where the transformation of $\tau_0$ is given by the following lemma


\begin{lemma}
\label{lemmatesch}
The function $\tau_0(\theta_0,\theta_1,\theta_\infty)$ satisfies 
\bea
\frac {\tau_0(\theta_0+1, \theta_1,\theta_\infty)}{\tau_0(\theta_0,\theta_1,\theta_\infty)} =   \frac{
 \Gamma(1-\theta_0 + \theta_1 + \theta_\infty) \Gamma(2\theta_0) \Gamma(2\theta_0-1)}
 {\Gamma(\theta_0 -\theta_1 + \theta_\infty) \Gamma(\theta_0 + \theta_1- \theta_\infty) \Gamma(\theta_0 + \theta_1 + \theta_\infty)}\;.
 \la{ST}
\eea
\end{lemma}
{\bf Proof.}
In order to apply \eqref{346} we need to find the matrix $Y_0(w)$. From \eqref{Psiw} we read off
\be
Y_0(w) = 
(1-w)^{\theta_1} \wt G_0^{-1}  H(w) \le[
 \begin{array}{cc}
 h_0(w) & w^{2\theta_0}s_0(w)\\\
 h_0'(w) &w^{2\theta_0} s_0'(w)
 \end{array}
 \ri] \a_0^{\frac {\s_3}2}
\ee
where $\wt G_0, H(w), h_0(w),  s_0(w)$ are in \eqref{Ghs} and $\a_0$ in \eqref{a0}. It is then a matter of explicit computation to verify the formula (\ref{ST}). 
\QED
Of course the Lemma \ref{lemmatesch} agrees with \eqref{DOZZ}, using the relation for the Barnes' $G$-function $G(z+1) = \Gamma(z) G(z)$.

%
%

\section{Discussion: zeros and poles of $\tau$}

According to the general results of Miwa \cite{Miwa}, Bolibruch \cite{Bol}, Malgrange \cite{Malg} and Palmer \cite{Palmer:Zeros} the tau-function is non-singular in the space of positions of singularities $\{t_j\}$ outside of diagonals $t_j=t_k$ for non-resonant residue matrices (in $2\times 2$ case this means that $2\theta_j\not \in \Z$). Zeroes of $\tau$ correspond to the locus where the inverse monodromy problem becomes unsolvable, or equivalently, an associated  vector bundle becomes holomorphically non-trivial, see \cite{Bol}. In the previous  analysis  the deformation space consisted only of the position of the poles; in the present case $N=3$  the dependence on $t_j$'s is essentially trivial and the divisor of   zeroes of $\tau$  lies in the space  of the monodromy data.  The general idea of analysis of singular locus of  $\tau$ in the space of monodromy data  could be  similarly based on  the analytic Fredholm theorem \cite{clancey}. Indeed, one could  analyze the singularity structure of the Malgrange form \eqref{malf} and show that its singularities are poles with positive integer residue along a co-dimension one subvariety, which would imply that the tau function has zeroes there. 

We can see   a glimpse of this general picture in our explicit formula  \eqref{DOZZ1}.
The Barnes' $G$-function has zeros at  $0, -1,-2,\dots$ and hence  $\tau$ has poles for 
$$\theta_j = -\frac 1 2, -1 , - \frac 3 2, \dots $$ 
for some $j=0,1,\infty$.
At this locus  the differential equation (\ref{eq}) is {\it resonant}.  Thus the occurrence of poles at these points is not in contradiction with the general wisdom that the tau function is locally analytic.

The zeros of the tau function (\ref{tauz1}) appear for $\vec \theta\in \mathfrak M$ where $\mathfrak M$ is the following  locus:
$$
\mathfrak M:= \Big\{  \theta_0+\theta_1+\theta_\infty\in -\N \Big\} \cup 
 \Big\{  - \theta_0+\theta_1+\theta_\infty\in -\N \Big\} 
$$
$$
 \cup  \Big\{ \theta_0-\theta_1+\theta_\infty\in -\N\Big\}
\cup 
 \Big\{  \theta_0-\theta_1+\theta_\infty\in -\N\Big\}\;.
$$

This locus $\mathfrak M$, that we can thus interpret as  the Malgrange divisor in the space of monodromy data, is  contained in the {\it reducible} locus, where the monodromy matrices (\ref{Mon0}), (\ref{Mon1}, (\ref{Monin}) generate a reducible group (either upper or lower triangular). This a singular locus in the character variety where the trace coordinates do not separate points. 

It is to be noticed, however, that the $\tau$ function is {\it not} a function on the character variety, namely it is not invariant under the action of conjugation of monodromies. The reason is that while the two--form $\omega_\Mcal$ \eqref{OM} is  well-defined on the space $\Mcal$,  the   potential $\Theta_\Mcal$ is not.  This explains the non-invariance of $\tau$ under the change of sign $\theta_j\to -\theta_j$.

{\bf Acknowledgements.}We thank J.Teschner for useful comments. 
The work of M.B. was supported in part by the Natural Sciences and Engineering Research Council of Canada (NSERC) grant RGPIN-2016-06660.
The work of D.K. was supported in part by the NSERC grant
RGPIN-2020-06816.

\bibliographystyle{MyBibStAlph}
\bibliography{Biblio.bib}

\end{document}